SOME INSIGHT
INTO
MANY CONSTITUENT DYNAMICS

by
Jerzy Hanckowiak

Zielona Góra University
ul. Podgórna 50, Poland, UE
e-mail: hanckowiak@wp.pl

ABSTRACT

A description of many constituent (particle) systems with fuzzy initial conditions is proposed with the help of the field language. In this language correlation functions are defined and equations for them are derived in the free Fock space. Additional conditions for their solutions are postulated.

The closure problem for polynomial, rational, local and nonlocal interactions is considered with exploitation of left or right invertibility of used operators.

Some general remarks concerning symmetry and multi-scale descriptions are given.

Key words:
Averages, generating vector, Cuntz relations, free Fock space, right and left inverse operators, local and nonlocal interaction, multi-scales

CONTENT

## 1. INTRODUCTION

A possible way to extend microscopic modelling and bridge it with experimental techniques is to do some averaging (coarse-graining). In general, a certain systematic method of integrating (averaging) of particularities occurring at a smaller scale may lead to effective and rational methods of description of more complicated systems, Rutgers (2004). The concept of hidden variables proposed by some founders of quantum mechanics and still developed by others, see 't Hooft (2001, 2006), also belongs to those kind of problems.

In other domains of human activities like geography or art we observe that passage from smaller scale to larger is accompanied by appearing of patterns and harmony. We say that some *emergent properties* appeared, Heller et al. (2006).

Considering many constituent systems we do not need to trace the exact "trajectories" of all constituents of a system. For example, to get the fast algorithms used in calculations, the particles which collide in the system as well as directions of their collisions can be randomly chosen. The model of hard spheres describes plenty of phenomena well. We believe that this is evidence that an interaction among micro-constituents can be substituted **by simple functions** which can be determined by means of macro experiments (phenomenological approach). This can be used for solving the closure problem for correlation functions or the moment problem in the replica method, Tanaka (2007).

For larger scale space-time processes described by the so called *effective equations* like Navier-Stokes equations and derived by means of a closure procedure, Dreyer et all (1999), further reduction of information about many-particle system achieved. It turns out however that even solutions to effective equations can depend sensitively on the initial and boundary conditions and extra averages are needed to get predictive results, Glimm et al. (1996). Even in slowly flowing dilute colloids, the particles "refuse to remain on purely parallel paths" and "favour chaos over calm", see "Picked up for you" (2008-01).

To correctly describe phenomena in a given scale, we need to find effective approximations for a few correlation functions or we need to effectively solve the closure problem for n-point correlation functions. Complicated perturbation approximations and lack of clear understanding of the closure mechanisms cause, however, that both these problems are only solved for a small or large values of the expansion parameters. To get some new insight into them, the free (full) Fock space was introduced. In such enlarged Fock space based on the Cuntz relations, many transformations can be executed and varies expansions for correlation functions can be proposed, see also Hanckowiak (2007). In the paper we also discuss varies aspects of the proposed formalism like a relation of symmetry of derived equations to symmetry of their solutions, a relation of number of variables to number of degrees of the freedom or an essence of difference between linear and nonlinear theories.

We start from the Newtonian equations. However, to pay a tribute to the great ancient philosopher, we describe the Newton's equations of motion in the Aristotelian form

$$dX / dt = F(X); \quad X \in R^{6N}$$

(1.1)

where $X = (\vec{X}, \vec{V}) = (\vec{x}_1, \vec{v}_1, ..., \vec{x}_n, \vec{v}_n) \equiv (x_1, ..., x_{6N})$ is a vector whose components can be the positions and velocities of particles or the concentration of species at a given time $t$, the corresponding components of $F(X)$ are the velocities and accelerations multiplied by inverses of masses of appropriate particles. We use uppercase letters for quantities related to the whole system and lowercase letters for individual particles.

## 1.1 LIOUVILLE AND BBGJK EQUATIONS

Even a good attempt at the investigation of Nature furnished by the Newton's equations does not apply in all circumstances. In the case of a system with large number of microscopic degrees of freedom it is not recommended to use a very precise and usually reversible description of the system following the Newton's description. In such cases, instead of 6N equations which Newton's approach offers to describe the dynamics of the system, we use only one equation. This can be an equation, which formally resembles us the equation upon an integral of motion for Eq. (1.1):

$$i\partial f(X;t) / \partial t = Lf(X;t); \quad f \geq 0$$

(1.1.1)

where

$$L = -iF(X) \cdot \nabla_X$$

(1.1.2)

is the Liouvillian, and the function $f$ defined on phase space of vectors $X$ is nonnegative and, for Hamilton systems, can be interpreted as a density of probability of finding the system in the state $X$. Eq. (1.1.1) with Hermitian operator $L$ is called the Liouville's equation.

In the Liouville's equation like in the Newton's equation, initial conditions do not enter explicitly and they are treated as random quantities. In the Newton's equation, the choice of a solution corresponds to unique choice of the positions and velocities of particles. In the Liouville's equation the choice of a solution corresponds to a unique density of probability of positions and velocities of entering particles (constituents).

On the other hand, for large systems (N-large), even Liouville's equations supply too much information from the point of view macroscopic description. Instead of distribution function $f(X;t)$ describing the whole system, we need distribution functions $f_n(t)$ describing only small parts of the system ($n << N$). We introduce

$$f_n(t) = (\Pi_n f)(t)$$

(1.1.3)

where $\Pi_m$ are projectors preserving, with growing $n$, more and more information of the original probability distribution $f(X,t)$:

the projectors $\Pi_m$ acting on the probability distribution $f$ are defined as follows:

$$(\Pi_m f)(x_{(m)};t) = V^{-(6N-m)} \int f(x_1, ..., x_m, x_{m+1}, ..., x_{6N};t) dx_{m+1}...dx_{6N} \equiv f_m$$

where $V$ is a volume of integration with respect to any – 1D variable $x$, Resibois et al. (1977). We have

$$\Pi_m \Pi_n = \Pi_m; \; m \le n$$

(1.1.5)

More general projectors on the n-point subsystems are considered by Toda et al. (1985). The functions $f_n$, (1.1.3-4), are called the *n-point partial probability distributions* or the *n-th marginals*, see Bardos (2000). To deal with infinite number of functions $f_n$ $(N \to \infty)$, the generating vector is introduced

$$\mid f > \Leftrightarrow \{1, f_1, ..., f_n, ...\}$$

(1.1.6)

where Dirac's notation for vectors is used. With the help of the generating vector |f> and projectors $\Pi_n$ one can describe an infinite system of equations upon partial distributions $f_n$ in a form of one, linear, evolution equation:

$$\partial_t \mid f > = L \mid f >$$

(1.1.7)

which are known as the Bogolubov-Born-Green-Jona-Kirkwood (BBGJK) chain of equations, Balescu (1975), Achiezer et al. (1977) or the Boltzmann hierarchy, Bardos (2000). In principle, this equation can describe a finite number of particles in a finite region and separated from the rest of the infinite number of particles by an infinite potential wall.

The Liouville operator $L$ acting in the linear space of generating vectors (1.1.6) has the following structure

$$L = L_D + L_{UD}$$

(1.1.8)

where the operator $L_D$ is a diagonal operator with respect to the projectors $P_n$ defined as follows:

$$P_n \mid f > = \mid f_n >$$

(1.1.9)

where the vectors $\mid f_n >$ represent partial probability distributions (1.1.4). This means that

$$P_n L_D = L_D P_n$$

(1.1.9)

The second operator in decomposition (1.1.8) is an upper diagonal operator with properties

$$P_n L_{UD} = P_n L_{UD} P_{n+1}$$

(1.1.11)

The last property is responsible for the *closure problem* in the world of marginals. A resignation from precise description of the system takes its revenge in impossibility of derivation of closed equations for marginals. In fact, to close an equation derived upon a given n-point function (n-pf), we have to say something about n+1-pf supplying a more precise description of the system ◆

## 1.2 "FIELD" NOTATION AND MULTITIME CORRELATION FUNCTIONS

In the paper, instead of the n-point partial probability distributions (marginals), we consider the averaged products of solutions to Eq. (1.1) called - the *multitime correlation* *functions* (MTCF). Similar quantities are used in the histories approach to quantum theory, Anastopoulos (2008). In the free Fock space we derive an equation for their generating vectors, |V>, analogues of (1.1.7-8). In this space we derive various formulas leading to different approximations of MTCF.

MTCF can be defined as follows:

$$< \Phi(\tilde{x}_1) \cdots \Phi(\tilde{x}_n) > \equiv \int \Phi[\tilde{x}_1; Y] \cdots \Phi[\tilde{x}_n; Y] \cdot f[Y] \cdot d^{6N}Y; \quad n = 1, 2, \ldots$$

$$(1.2.1)$$

where the "field" $\Phi[\tilde{x}; Y]$ describes some properties of the considered dynamical system (1.1) and will be interpreted below. The components of $\tilde{x}$, which fully characterize the type and position of the particle or density of constituent, can be discrete and continuous. It also contains the time. Due to such concentration of information, it is easy to see the permutation symmetry of the considered correlation functions. In other words, as in Faraday's times the notion of field still plays a role of a simplification factor. The function(al) $f[Y]$ can be related to the initial probability distribution:

$$f[Y] \Leftrightarrow f(X; 0)$$

$$(1.2.2)$$

The "field" $\Phi(\tilde{x}; Y)$ in the formula (1.2.1) was introduced to describe in a *universal way* the various methods of description of systems. For example, for

$$\tilde{x} = (\alpha, j, t), \quad where \quad \alpha = 1, 2, 3, \quad j = 1, 2, \ldots, N$$

$$(1.2.3)$$

$$\Phi[\tilde{x}; Y] \equiv x_{\alpha j}(t; Y) \quad and \quad \Phi[\alpha, j, 0; Y] = y_{\alpha j}$$

$$(1.2.4)$$

describes the $\alpha$-th component of the radius vector of the j-th particle with the all initial conditions described by the vector $Y$. For a continues system, $\tilde{x} = (\alpha, \vec{x}, t) \equiv (\alpha, x)$, where the vector $\vec{x}$, instead of the previous discrete index $j$, identifies points of the system (Lagrange or material description). We can also use a description in which the variable $\vec{x}$ denotes a point in the space (Euler). The field $\Phi$ can also be interpreted as displacements from equilibrium in the case of lattice systems. For another exploitation of the field concept, see Gomis et al. (2000) and Bering (2000).

We also use the notation:

$$\Phi(\tilde{x}) \equiv \Phi(\alpha, x) \equiv \Phi(\alpha, \chi, \vec{x}, t) \Leftrightarrow \Phi[\tilde{x}; Y]$$

$$(1.2.5a)$$

with

$$\tilde{x} = (\alpha, x) = (\alpha, \chi, \vec{x}, t)$$

$$(1.2.5b)$$

in which a particle is identified by means of two symbols: the $\vec{x}$ (its initial position) and the letter $\chi$ describing the type of particle. So, for invariants, we use the following notation

$$\Phi^2(x) = \Phi^2(\chi, \vec{x}, t) = \sum_\alpha \Phi^2(\alpha, \chi, \vec{x}, t), \quad \Phi^2(\chi, t) = \sum_\alpha \int d\vec{x} \Phi^2(\alpha, \chi, \vec{x}, t); \quad x = (\chi, \vec{x}, t)$$

$$(1.2.5c)$$

For $t \neq 0$, $\Phi$ may describe the position of a particle at the time $t$.

It turns out that with averages (1.2.1), an additional spatial and/or temporal integration can be introduced:

$$< \Phi(\tilde{x}_1) \cdots \Phi(\tilde{x}_n) >\equiv$$

$$\int \Phi[\alpha_1, x_1 - w; Y] \cdots \Phi[\alpha_n, x_n - w; Y] \cdot f[Y] \cdot W(w) \cdot dY \cdot dw$$

$$(1.2.7)$$

These double averages can be used when we do not control the initial conditions together with identification of particles (Lagrange), or points in space (Euler), and instances of measurements. One can also connect one kind of averaging with a chosen scale and other type with a measurement precision of the used instruments.

It is easily seen that, for the translation invariant theories and other symmetries, both kinds of averages considered separately or simultaneously, like in (1.2.7), lead to identical equations for correlation functions.

By averaging, we do not diminish the number of variables but we consider more and more smooth and similarly behaving variables $< \Phi(\tilde{x}) >; \tilde{x} \in R^{d+1}$ what is equivalent with the more *coarse-grained description of the system*. So, instead of looking for a few, essential macroscopic variables, we consider the same number of averaged microscopic variables with more mild behaviour and clear interpretation. Such an abundance of variables together with the idea of full Fock space allows us to describe in a more uniform way the multi-scale phenomena. In fact, the used averaging procedures remind us the mathematical procedure of abstraction based on some equivalent relation $\rho$ by means of which in a given set $S$ classes of equivalent elements [s], where $s \in S$, are constructed. The number of equivalent class is exactly such as a number of elements of the set $S$. However, decreases the number of different equivalent class and this process is more intensive for more large equivalent classes [s].

The most important difference between BBGJK Eq. (1.1.7) and equations derived for MTCF, called in I the Reynolds-Kraichnan-Lewis equations (RKLE), and considered below, is that in the latter equation the terms describing linear interactions enter a diagonal part of the operator describing RKLE. In the case of (1.1.7) - both, linear and non-linear terms of interaction enter the upper diagonal part of the operator (1.1.8) which is responsible for the closure problem, see (1.1.11). This means that even linear systems, e.g. a chain of linear oscillators, lead to an infinite chain of branching BBGJK equations of type (1.1.7).

### 1.3 SYMMETRY

Very often, the derived equations exhibit certain kind of symmetry, which are not sheared by their physical solutions. In this way the spontaneously broken symmetry (SBS) idea, exemplified in Quantum Field Theories (QFT), arises, Jackive (1996). It is interesting that to eliminate the divergences accompanying to every local QFT (additional condition), SBS arises. In fact, SBS accompanies also to classical theories described by Eq.(1.1) and is introduced by the initial conditions; gravitation interaction is spherically symmetrical but Kepler's laws not, Jackiev (1996). SBS is also important in solid state physics considering crystal with appropriate symmetry. According to the Goldstone theorem, vibrations with macroscopic (low) frequencies are possible due to SBS.

In this paper we have quite opposite situation concerning a (permutation) symmetry. First, we introduced correlation functions $< \Phi(\tilde{x}_1)...\Phi(\tilde{x}_n) >$ which are permutation symmetrical with respect to changing of variables containing also information about type of particles or constituents. Derived equations for correlation functions, however, are not symmetrical at all (one variable ($\tilde{x}_1$) is distinguished) even so they admit symmetrical and non-symmetrical solutions. Next, we introduce the full Fock space (free Fock space) encompassing symmetrical (physical) and non symmetrical (unphysical) correlation functions what allows us for appropriate transformations of derived equations for correlation functions. The fact that we are interested in symmetrical solutions is next used in the ordering problem in the vector description of considered equations, Sec.3.

A redundancy in the description of a system is commonly used in theoretical physics, for example in gauge theories, where in this way a connection between the free and the interaction parts of a theory are introduced. In our opinion we have similar redundancy in the case of equations for correlation functions, but it is lost by too fast introduction of symmetrical generating vectors. In this way, without any reasons, a privilege role of the free part of theory is exposed.

The theory based on the n-point functions invariant with respect to changing of particles of different kinds means the super-symmetry. This is obtained due to an appropriate extension of field language and used averages.

## 2. GENERAL FRAME OF DESCRIPTION

Ideally we would like to know how much we can simplify interaction because of used averages and the fact that we consider a big number of particles. To get a first insight into this problem we introduce a whole family of forces which depend on n-point functions with more or less direct interpretation and we would like to obtain a general solution to considered equations for correlation functions.

### 2.1 DYNAMIC EQUATIONS

Let us consider a system which dynamical properties are described by the "field" $\Phi(\tilde{x}, t)$, see (1.2.3-5), with an appropriate interpretation (e.g., radius or displacement vector or constituent density). We assume that the field $\Phi$ satisfies the following forminvariant (Lopuszanski (1998; page 125)) dynamical equation

$$K\Phi(\tilde{x}) = \Phi(\tilde{x}) \cdot M[\Phi; x] + G(\tilde{x}); \quad \tilde{x} = (\alpha, x); \quad x = (\chi, \vec{x}, x_0) \equiv (\chi, \vec{x}, t)$$

(2.1.1a)

or

$$K\Phi_\alpha(x) = \Phi_\alpha(x) \cdot M[\Phi; x] + G_\alpha(x)$$

(2.1.1b)

where $K$ is a linear operator including the second derivative with respect to time $t$ (Newton's equations) and the scalar functional $M$ describes a nonlinear interaction of particle $P$ with itself and other constituents of the system. $G$ describes an external force acting on the particle $P \Leftrightarrow (\chi, \vec{x})$. We assume that the scalar functional $M$ has the form

$$M[\Phi; x] = \sum_n \tilde{M}_n(\tilde{z}_1, ..., \tilde{z}_n; x)\Phi(\tilde{z}_1)...\Phi(\tilde{z}_n)d\tilde{z}_{(n)} =$$

$$const + \sum_{n=1} \int M_{2n}(\tilde{x}_1, \tilde{y}_1, ..., \tilde{x}_n, \tilde{y}_n; x)\big(\Phi(\tilde{x}_1) - \Phi(\tilde{y}_1)\big)...\big(\Phi(\tilde{x}_n) - \Phi(\tilde{y}_n)\big)d\tilde{x}_{(n)}d\tilde{y}_{(n)}$$

(2.1.2)

where the 2n-point functions (2n-pfs) $M_{2n}(\tilde{x}_{(n)}, \tilde{y}_{(n)}; x)$ describe in general, a variety of non-additive interactions including equations with transformed arguments, Przeworska-Rolewicz (1973).

*Additive interaction* is described by the scalar (invariant) functional

$$M[\Phi;x] = \sum_n \int M_{2n}(y;x)(\Phi(x)-\Phi(y))^{2n}dy$$

(2.1.3)

constructed by means of invariants (scalars)

$$\left(\Phi(x)-\Phi(y)\right)^2 \equiv \sum_{\alpha=1}^{3}\left(\Phi(\alpha,x)-\Phi(\alpha,y)\right)^2$$

(2.1.4)

The index $n$ in the above sum runs over entire or/and rational numbers. The case of functions $M_{2n}$ not depending on $\vec{x}$ and $\vec{y}$ means that interaction of every particle with every other particle is taken into account with the same weight.

The case

$$M_{2n}(\tilde{y};\tilde{x}) = M_{2n}(y;x) = \delta(t'-t)M_{2n}(\chi',\vec{y};\chi,\vec{x})$$

(2.1.5)

corresponds to Newton's action at a distance and leads to the time translation symmetry of the system.

The case

$$M_{2n}(\tilde{y};\tilde{x}) = M_{2n}(y;x) = \delta(t'-t)M_{2n}(\chi',\chi,\vec{y}-\vec{x})$$

(2.1.6)

is more demanding. It leads, with additional conditions, to space-time translation symmetry of solutions. In other words, if $\Phi(\alpha,\chi,\vec{x},t)$ satisfies Eq. (2.1.1) then also $\Phi(\alpha,\chi,\vec{x}+\vec{w},t+t_0)$ satisfies Eq. (2.1.1). These conditions are fulfilled in the case of space-time translation invariance of Eq. (2.1.1). For discrete systems (an infinite collection of particles), the vectors $\vec{x}$ and $\vec{w}$ are discrete and the above "space" translation invariance, even natural, imposes a restriction upon the linear term in Eq. (2.1.1).

It is interesting that for lattice systems the long-range interaction like the Coulomb interaction is expressed by the linear terms but the case of short-range interaction between neighbours is expressed by the nonlinear terms, see e.g. Mingaleev et al. (1998), Eleftheriou et al. (2000) and Laskin et al. (2006).

## 2.2 ONE-PARAMETER FAMILY OF INTERACTIONS; Q-DEFORMATION

Now, let us introduce one parameter family of functionals

$$M_q[\Phi;x] = M_q[\Phi;P,t] =$$
$$\sum_n \int M_n(\tilde{x}_1,\tilde{y}_1,...,\tilde{x}_n,\tilde{y}_n;x)(\Phi(\tilde{x}_1)-q\Phi(\tilde{y}_1))..(\Phi(\tilde{x}_n)-q\Phi(\tilde{y}_n))d\tilde{x}_{(n)}d\tilde{y}_{(n)}$$

(2.2.1)

with an arbitrary *deformation parameter q*. We can look on this formula as a q-deformed formula (2.1.2) or parameter $q$ can be treated as a *coupling constant of the non-local theory*, see also Gomis at al. (2004). $q$-deformed formula (2.1.3) is

$$M_q[\Phi;x] = M_q[\Phi;P,t] = \sum_n \int M_{2n}(y;x)(\Phi(x)-q\Phi(y))^{2n}dy$$

(2.2.2)

It describes *deformed additive interactions*. For $q=0$, we have self-interaction. For $q=1$, we have initial interactions of particles. It is also possible that $q \neq 1$ can be interpreted as a parameter describing screening of the original interaction of a single pair of particles caused by many body situation.

## 2.3 FINITE NUMBER TERMS HYPOTHESIS

To further specify an interaction among particles, in the case of a large number N, let us assume that functions

$$M_n(\tilde{x}_1, \tilde{y}_1, ; , \dots; \tilde{x}_n, \tilde{y}_n; x) \neq 0$$

(2.3.1)

in (2.1.2) or (2.2.1) or (2.2.2) - **only for a finite numbe**r of $n$, see Mingaleev et al. (1998), Eleftheriou et al. (2000).

The above assumption is justified by classical or quantum physics, see Newton's equations in which the force is given by a sum of the same order monomial terms, or in various kinds of quantum field theories with interaction terms described most often by compact formulas. In the case of many constituent systems one can expect additional simplification of interaction terms, see Introduction. As a result, effective interactions can be described by more useful functions than the original ones. For a many particle system, stable composite objects can exist even for repulsive interactions, Winkler et al. (2006). It is also known that the certain class of nonlinear systems can be embedded in a bilinear system of higher dimensionality, Phillips (2000), or in the case of biochemical reactions, for which most constituent reactions are between only two or three species at a time, see Bamieh et al. (2006).

What interaction between particles are assumed when the Navier-Stokes equation (NSE) is derived? It is surprising how little specific shape of the interaction influences the final NSE derived by means of the Boltzmann equation (BE), see Huang (1963, (1978; pages 98, 109)). See also Dorfman (1998), where to derive BE, the repulsive forces between particles are only assumed. More important are conservation laws, which are satisfied for considered type of interactions.

## 2.4 A HYBRID DESCRIPTION

Now we consider MTCF:

$$< \Phi(\tilde{x}_1) \cdots \Phi(\tilde{x}_n) > = \int \Phi[\tilde{x}_1 - \tilde{w}; Y] \cdots \Phi[\tilde{x}_n - \tilde{w}; Y] W(\tilde{w}) \, d\tilde{w} \cdot f[Y] \, dY; \quad n = 1, 2, \dots$$

(2.4.1)

where $\Phi[\tilde{x}; Y]$ represents the general solution to Eq. (2.1.1) with $Y$ representing initial conditions imposed upon the system. These initial vectors are random variables which distribution function $f[Y]$ should be chosen according to rules concerning macroscopically definable ensembles, Gorban et. all (2003). In this description - variable $\tilde{x} = (\alpha, \chi, \vec{x}, t)$ - describes internal as well as space-time characteristic of a particle. An integration, or rather summation, with respect to variable $\alpha$ can be interpreted as averaging with respect to different reference of frames.

Let us consider a *hybrid description* in which it is claimed that some, usually small part of the body is described by means of the molecular dynamics (MD) method and the rest of the body – by the field dynamic (FD) method, Liu (2004). Let us designate points described by MD method by position vectors $(\vec{y}_1, \dots, \vec{y}_k)$ and points described by FD method by $(\vec{y}_{k+1}, \dots, \vec{y}_N)$. In the orthodoxies FD, the points $(\vec{y}_{k+1}, \dots, \vec{y}_N)$, in fact, are only a representation of the continuous medium. A different descriptions of these two groups of points can be expressed by the function (functional) $f$. We assume that first group of points treated by MD has much more precise initial values then the second group of points. So, we assume that

$$f \propto \delta(\tilde{y}_1 - \tilde{y}_{01})\delta(\tilde{v}_1 - \tilde{v}_{01})...\delta(\tilde{y}_k - \tilde{y}_{0k})\delta(\tilde{v}_k - \tilde{v}_{0k})g(\tilde{y}_{k+1}, \tilde{v}_{k+1},...,\tilde{y}_N, \tilde{v}_N)$$

(2.4.2)

where the function *g* has a support corresponding to a precision of the FD.

In fact, all times we have the same number of variables describing the system, but in the field area they correspond to much more smeared initial conditions. More important, however, is here fact that, with introduction of MTCF, we do not need to use the non-linear original Eq.(2.1.1) or (1.1) in any area of the body. These equations are substituted by the system of linear equations for MTCF considered in Sec.3. This allows us to use the linear operator constructions to solve them. Moreover, the correlation functions, including the 1-point function, with exception of small areas, are smooth quantities which additionally simplify their construction.

Because with the help of appropriate choice of the density probability *f* and smearing function *W* we get exact (not averaged) solutions to Eq. (2.1.1), we can look on MTCF, (2.4.1), and derived for them Eqs (3.2.1) as on some version of the Carleman linearization technique applied to Eq. (2.1.1), see Kowalski et al. (1991) and Kowalski (1998). A main difference with the original Carleman technique consists in that our new "variables", MTCF, needed for linearization of the original Eq (2.1.1), are rather related to averaged solutions and their products than to exact solutions.

## 2.5 DIFFERENCE BETWEEN LINEAR AND NONLINEAR THEORIES

It is also interesting to stress here a radical difference between the linear and non-linear theories (2.1). For the linear theories, smooth (coarse) and non-smooth (fine variables) (1-pf) may satisfy exactly the same – linear equations. In such theories, the passage from fine to coarse variables takes place exclusively by means of the initial and boundary conditions. A situation is different when the linear theories are described by inhomogeneous equations. In this case, the time-space averages are not equivalent with the initial condition averages. Similarly, in the case of inhomogeneous linear equations the smooth Wigner transforms used for effective numerical calculations lead to other equations, see Athanassoulis et al. (2008).

A non-linear term in Eq.(2.1) is responsible for less or more complicated relations between fine and coarse variables and leads to the closure problem in equations for MSCF. In general, we can say that relations between $\mu$-world and *M*-world are coded in the nonlinear terms of considered dynamic equations.

There are many examples in science, in which nonlinearity always essentially differs from the linear case, see Chiu (2004), Scott (2007).

We consider now simple examples of Eq.(2.1) and analyse the meaning of (2.4.2). First, let us consider the 1D wave equation

$$\left(\partial_t^2 - \alpha^2 \partial_x^2\right)\Phi(x,t) = 0$$

(2.5.1)

Solutions satisfying homogenous boundary condition are given by the d'Alembert formula

$$\Phi(x,t) = \frac{\Phi(x+\alpha t,0) + \Phi(x-\alpha t,0)}{2} + \frac{1}{2\alpha}\int_{x-\alpha t}^{x+\alpha t}\dot{\Phi}(s,0)ds$$

(2.5.2)

with the arbitrary initial conditions

$$\left(\Phi(x,0),\dot{\Phi}(x,0)\right)\big|_{x\in B} \leftrightarrow Y$$

(2.5.3)

where $B$ is a set describing the system. The formula (2.5.2) shows an influence of the initial displacements $(\Phi(x,0))|_{x \in B}$ and velocities $(\dot{\Phi}(x,0))|_{x \in B}$ of particles (points) on the temporal position $(\Phi(x,t))|_{x \in B}$ of the particle $x$. We see that even for the linear systems, MD and FD regions, used simultaneously, influence each other; there are no enclaves described exclusively by, e.g., MD. Let us now assume like in (2.4.2) that the $k$ "initial points" are exactly determined and the rest of the initial points are random Gaussians with vanishing averages. This means that instead of one system we consider an ensemble of systems described by the probability distribution (2.4.2). In this ensemble the $k$ particles have exact initial dates. For the linear systems, the averages satisfy the same equations as the original quantities. In the case of (2.4.1-2), we have

$$< \Phi(x,t) > = \frac{< \Phi(x+\alpha t,0) > + < \Phi(x-\alpha t,0) >}{2} + \frac{1}{2\alpha} \int_{x-\alpha t}^{x+\alpha t} < \dot{\Phi}(s,0) > ds$$

(2.5.4)

The distribution (2.4.2) means that the measurements, with running time, will repeat at different collections of points giving, after averaging, zeros displacements at complementary regions (e.g. Gauss distributions). This picture is quite natural if measures are substituted by our eyes. We will see only displacements of some permanent features.

The above remarks concerned an influence of initial positions of points on the averaged evolution of the system. An influence of initial velocities of points, described by third term of (2.5.4), is less significant because of the integration interval - linearly growing in time $t$.

## 3. EQUATIONS FOR MULTI-TIME CORRELATION FUNCTIONS (MTCF); REYNOLDS-KRAICHNAN-LEWIS TYPE EQUATIONS (RKLE).

### 3.1 FREE (FULL) FOCK SPACE; CUNTZ (CO-) RELATIONS

MTCF (*moments*) can be handled with the help of generating vector |V>, also considered in I (Hanckowiak (2005, 2007)),

$$|V> = \sum_n \int < \Phi(\tilde{x}_1) \cdots \Phi(\tilde{x}_n) > \hat{\eta}*(\tilde{x}_1) \cdots \hat{\eta}*(\tilde{x}_n) | 0 > d\tilde{x}_{(n)}$$

(3.1.1)

Orthogonal vectors $\hat{\eta}*(x_1) \cdots \hat{\eta}*(x_n) | 0 >$ are constructed by means of the operators $\hat{\eta}*$ satisfying the Cuntz (co-) relations

$$\hat{\eta}(\tilde{x}) \cdot \hat{\eta}*(\tilde{y}) = \delta(\tilde{x} - \tilde{y}) \cdot \hat{I}$$

(3.1.2)

They can be treated as generators of the $C^*$-algebra. The $\delta(\tilde{x} - \tilde{y})$ is an appropriate dimensional Dirac's delta. In the case of discrete components of the vectors $\tilde{x}, \tilde{y}$, the Dirac's deltas are substituted by the Kronecker's deltas and integrations are substituted by summations. $\hat{I}$ is the unit operator. Cuntz relations (3.1.2) are appropriate for polynomial and rational interactions. In literature are also considered other Cuntz relations, see e.g., Halpern at al. (1999). In fact, instead of vectors (3.1.1) we could use linear functionals

$$V = \sum_n \int < \Phi(\tilde{x}_1) \cdots \Phi(\tilde{x}_n) > \alpha(\tilde{x}_{(n)}) d\tilde{x}_{(n)}$$

with "arbitrary" n-point functions $\alpha(\tilde{x}_{(n)})$. Cuntz approach, however, guarantees a more compact description of appearing operators.

We assume also that "vacuum" bra and ket vectors exist with properties

$$\hat{\eta}(\tilde{x}) \mid 0 >= 0, \quad < 0 \mid \hat{\eta}*(\tilde{x}) = 0; \quad (\tilde{x}) \in R^{d+1}$$

(3.1.3)

Using the quantum terminology, the auxiliary operator field $\hat{\eta}$ "annihilates" the "vacuum" vector |0>. In other words, we assume the *free Fock space F* made of the vectors (3.1.1). Adjective *free* is used here to stress that basic operator variables $\hat{\eta}, \hat{\eta}*$ are free from additional restrictions on commutation relations. It is also interesting to compare the above construction of the free Fock space with other based on the tensor products of the Hilbert space *H* known under name the *interacting free Fock space*, Lu (1997).

For a representation of the Cuntz relation in the finite case, see Jorgensen (1999) and in the infinite case, see Aref'eva et al. (1994).

With the help of relations (3.1.2-3) one can easy prove the following formula for MTCF:

$$< 0 \mid \hat{\eta}(\tilde{x}_1) \cdots \hat{\eta}(\tilde{x}_n) \mid V >=< \Phi(\tilde{x}_1) \cdots \Phi(\tilde{x}_n) >$$

(3.1.4)

We see that the above scalar products of the generating vector $\mid V >$ with the vectors $< 0 \mid \hat{\eta}(x_1) \cdots \hat{\eta}(x_n)$ can be interpreted as averaged solutions and correlation functions of Eq. (2.1.1). By means of these quantities one can build physical macroscopic quantities, see Sec.VIII.

We have used the Dirac's notation in which the vertical bar means the beginning of the *ket vector* |V> or the end of the *bra vector* <V|. In the product of a bra $< \Psi \mid$ and a ket vectors $\mid V >$ only one vertical bar is left: $< \Psi \mid V >$

## 3.2 EQUATIONS FOR GENERATING VECTOR OF CORRELATION FUNCTIONS. NORMAL ORDERING

With these tools and Eq.(2.1.1) one can derive the following equation for the generating vector (3.1.1):

$$\left( \hat{K} + \hat{N} + \hat{G} \right) \mid V >= 0$$

(3.2.1)

with a diagonal operator

$$\hat{K} = \int \hat{\eta}*(\tilde{x}) K(\tilde{x}, \tilde{y}) \hat{\eta}(\tilde{y}) \, d\tilde{x} d\tilde{y}$$

(3.2.2)

related to the linear part of Eq.(2.1.1), see (3.3.1) and below. An upper triangular operator

$$\hat{N} = \int \hat{\eta}*(\tilde{z}) \hat{\eta}(\tilde{z}) : M[\hat{\eta}; z] : dz$$

(3.2.3)

is related to the non-linear term of Eq.(2.1).

Sign :...:, called the normal ordering, means that the order of operators $\hat{\eta}(\tilde{x})$ in (3.2.3) is not important. This is because the operators $\hat{\eta}$ of the operator $\hat{N}$ occurring in Eq. (3.2.1) act on permutation symmetrized products of the operators $\hat{\eta}*$ appearing in the expansion (3.1.1) (From definition (1.2.1) or (1.2.7) correlation functions $< \Phi(\tilde{x}_1)...\Phi(\tilde{x}_n) >$ are permutation symmetric with respect to variables $\tilde{x}$. This leads to symmetrization of the products of the operators $\hat{\eta}*$ in the integrals or rather sums of (3.1.1), see Rzewuski (1969; page 21).

The lower triangular operator

$$\hat{G} = \int \hat{\eta}*(\tilde{x})G(\tilde{x})d\tilde{x}$$

(3.2.4)

is related to the external force $G$ acting on the particles identified by the variables $\tilde{x}$. The external force can be also interpreted as an input to the nonlinear system and the presented paper as an alternative approach to the model reduction problem, see Phillips (2000), where it is written: "*Model reduction* also refers to the procedure of automatic generation of system macromodels from detailed descriptions. These macromodels can be used to perform rapid system-level simulation of engineering designs that are too complicated to analyse at the detailed component level".

It is also remarkable that these projection properties of Eq.(3.2.1) with respect to projectors (3.3.2) are very similar to projection properties of equations for the Green's functions of Quantum Field Theory (Schwinger equations), see Hanckowiak (1979). However, with the help of operators satisfying the co-Cuntz relations (3.1.2) all operators entering Eq.(3.2.1) are simplified. For comparison, see Rzewuski (1969), Hanckowiak (1992).

In all these formulas we tacitly assume that an integration is substituted by a summation when variables take discrete values.

## 3.3 COMPLETE SET OF PROJECTORS. ALGEBRA OF OPERATORS

Due to the presence of the operator $\hat{N}$ or $\hat{N}_q$, Eq.(3.2.1) generates an infinite chain of equations for n-pfs $< \Phi(\tilde{x}_1)...\Phi(\tilde{x}_n) >$. These operators are upper triangular operators with respect to the Hermitian projectors $\hat{P}_n$ defined as

$$\hat{P}_n \mid V > = \int d\tilde{x}_{(n)} < \Phi(\tilde{x}_1)...\Phi(\tilde{x}_n) > \hat{\eta}*(\tilde{x}_1)...\hat{\eta}*(\tilde{x}_n) \mid 0 >$$

(3.3.1)

The projectors $\hat{P}_n$ can be expressed by means of the operators $\hat{\eta}, \hat{\eta}*$ satisfying relations (3.1.2) using the Kronecker product of the bra and ket basis vectors (outer or tensor product):

$$\hat{P}_n = \int d\tilde{x}_{(n)} \hat{\eta}*(\tilde{x}_1)...\hat{\eta}*(\tilde{x}_n) \mid 0 > < 0 \mid \hat{\eta}(\tilde{x}_n)...\hat{\eta}(\tilde{x}_1)$$

(3.3.2)

see Wikipedia , <bra – ket notations>. In a representation in which * can be interpreted as the Hermitian conjugation, these are Hermitian projectors satisfying the orthonormality relations

$$\hat{P}_m \hat{P}_n = \delta_{mn} \hat{P}_n$$

(3.3.3)

and the completeness equality

$$\sum_{n=0} \hat{P}_n = \hat{I}$$

(3.3.4)

with $\hat{P}_0 = \mid 0 > < 0 \mid$.

## 3.4 *-ALGEBRAIC QUANTUM PROBABILITY SPACE

A general type of linear operators acting in Fock space of generating vectors (3.1.1) can be presented as follows:

$$\hat{A} = \sum_{m,n} \int dx_{(m)} dy_{(n)} A(x_{(m)}, y_{(n)}) \hat{\eta}*(x_1)..\hat{\eta}*(x_m) \hat{\eta}(y_1)...\hat{\eta}(y_n) + a \mid 0 > < 0 \mid$$

(3.4.1)

They form a formal $*$-algebra $A$ with multiplication and addition and with certain n-point operators $\hat{A}(\tilde{x}_{(n)})$ which "vacuum" expectation values

$$< 0 \mid \hat{A}(\tilde{x}_{(n)}) \mid 0 > = < \Phi(\tilde{x}_1) \cdots \Phi(\tilde{x}_n) > = V(\tilde{x}_{(n)})$$

(3.4.2)

have interpretation of MTCF, see (1.2.1). See also Sec.4. The last operation can be treated as a functional $\phi$ which together with the above $*$-algebra form $*$-*algebraic quantum probability space*. This is unital $*$-algebraic quantum probability space because the unit operator

$$\int d\tilde{x} \hat{\eta}*(\tilde{x}) \cdot \hat{\eta}(\tilde{x}) + \mid 0 > < 0 \mid = \hat{I}$$

(3.4.3)

belongs to the set constituted from operators (3.4.1). A *state* $\mathbf{P}$ on $A$ is a linear map $\mathbf{P}$ : $A \to C$ such that $0 \le \mathbf{P}(A*A)$ for all A from $A$ (positivity) and $\mathbf{P}(\hat{I}) = 1$ (normalization), see Wikipedia (Quantum probability). It is easy to see that we can choose:

$$< 0 \mid \bullet \mid 0 > = \mathbf{P}(\bullet)$$

(3.4.4)

So, in our case, the $*$-algebraic quantum probability space is the pair $(A, <0| \bullet |0>)$.

The last term in (3.4.1) represents the tensor product of two vectors $\mid 0 >$ and $<0 \mid$ and cannot be constructed as a product of operators $\hat{\eta}*, \hat{\eta}$. It is also interesting to notice that for operators constructed from the tensor products of basic vectors $\hat{\eta}*(x_1)..\hat{\eta}*(x_n) \mid 0 >$ :

$$\hat{A} = \sum_{m,n} \int dx_{(m)} dy_{(n)} A(x_{(m)}, y_{(n)}) \hat{\eta}*(x_1)..\hat{\eta}*(x_m) \mid 0 > < 0 \mid \hat{\eta}(y_1)...\hat{\eta}(y_n) + a \mid 0 > < 0 \mid$$

(3.4.5)

action of such an operator on a vector |V> can be described in usual matrix-vector notation.

# 4. VARIES EXPANSIONS OF GENERATING VECTORS. CHOICE OF THE ARBITRARY ELEMENTS

In many physical and technical situations the linear problems are formulated in such a way that a source of the unknown field, or an input, may take an arbitrary form. This is the main reason that corresponding operators used for descriptions of such problems are right invertible, see the Poisson, Maxwell or Navier-Stokes equations. But a more important feature of a right invertible formulation of a given problem or rather a class of problems is that many solutions associated to a given source (field) may exist. In the case of the source-free solutions, these, so called homogenous solutions, have often independent interpretations - like electromagnetic waves in case of Maxwell equations.

With the help of right inverse operators constructed in the free Fock space one can create various expansions for the generating vectors $|V>$ and construct appropriate projectors describing undetermined elements, Przeworska-Rolewicz (1989). In the free Fock space, however, there is the "initial" value problem which reminds us in some way the initial value problem of the dynamics with infinitely many derivatives, Barnaby et al. (2008). Subsection (4.1) and Sec.7 show us a possible remedy for that problem.

## 4.1 CANONICAL PERTURBATION THEORY. CHOICE OF ARBITRARY PROJECTIONS

To be more concrete, let us assume that the operator $\hat{K} + \hat{G}$ of Eq.(3.2.1) is a right invertible in the subspace $(\hat{I} - \hat{P}_0)F$. This means that an operator $(\hat{K} + \hat{G})_R^{-1}$ exists that

$$(\hat{K} + \hat{G})(\hat{K} + \hat{G})_R^{-1} = \hat{I} - \hat{P}_0$$

(4.1.1)

The projector (oblique projector)

$$\hat{P}_{K+G} \equiv \hat{I} - (\hat{K} + \hat{G})_R^{-1}(\hat{K} + \hat{G}) \equiv \hat{I} - \hat{Q}_{K+G} \quad with \quad \hat{P}_0\hat{P}_{K+G} = \hat{P}_0$$

(4.1.2)

projects on the null space of the operator $\hat{K} + \hat{G}$. With the help of these operators, RKLE (3.2.1) can equivalently be describe as follows

$$\left(\hat{I} + (\hat{K} + \hat{G})_R^{-1}\hat{N}\right)|V> = \hat{P}_{K+G}|V>$$

(4.1.3)

with an arbitrary projection $\hat{P}_{K+G}|V>$. For a construction of a right inverse operator $(\hat{K} + \hat{G})_R^{-1}$, see Sec.5. This means that if the inverse to the operator $\left(\hat{I} + (\hat{K} + \hat{G})_R^{-1}\hat{N}\right)$ exists then a general solution to Eq.(3.2.1) or equivalent Eq. (4.1.3), in the space $F$, can be represented as follows

$$|V> = \left(\hat{I} + (\hat{K} + \hat{G})_R^{-1}\hat{N}\right)^{-1}\hat{P}_{K+G}|V> = \sum_j (-1)^j [(\hat{K} + \hat{G})_R^{-1}\hat{N}]^j \hat{P}_{K+G}|V>$$

(4.1.4)

with an arbitrary projection $\hat{P}_{K+G}|V>$.

The vector $\hat{P}_{K+G}|V>$ can be constructed directly from definitions (1.2.1) or (1.2.7), on the assumption that $\hat{N} = 0$. In this case

$$\hat{P}_{K+G}|V> = \hat{P}_{K+G}|V>^{(0)} = |V>^{(0)} = \hat{S}|V>^{(0)}$$

(4.1.5)

In perturbation language, the assumption (4.1.5) means that the linear part of the theory, described by the operator $\hat{K} + \hat{G}$, does not depend on the perturbation (nonlinear) part described by the operator $\hat{N}$. The assumption (4.1.5) expresses long time lasting fundamental assumption about separation of the linear and nonlinear phenomena. However, the above choice may lead to non-symmetrical correlation functions. So, we substitute the assumption (4.1.5) by another one:

$$\hat{P}_{K+G} \mid V >= \hat{P}_{K+G}(\mid V >^{(0)} + additional \ terms) \Leftrightarrow \mid V >= \hat{S}(\hat{I} + (\hat{K} + \hat{G})_R^{-1} \hat{N})^{-1} \mid V >^{(0)}$$

(4.1.6)

where the projector $\hat{S}$ projects upon the permutation symmetric n-point functions, see I, Sec.4.

## 4.2 A NEW EXPANSION

Assuming that the operator $\hat{N}$ is a right invertible operator and hence

$$\hat{N}\hat{N}_R^{-1} = \hat{I} - \hat{P}_0$$

(4.2.1)

we can express the general solution to Eq.(3.2.1) in a similar way as in (4.1.4):

$$\mid V >= \left(\hat{I} + \hat{N}_R^{-1}[\hat{K} + \hat{G}]\right)^{-1} \hat{P}_N \mid V >= \sum_n (-1)^n \left(\hat{N}_R^{-1}[\hat{K} + \hat{G}]\right)^n \hat{P}_N \mid V >$$

(4.2.2)

where the projector

$$\hat{P}_N = \hat{I} - \hat{N}_R^{-1}\hat{N} \quad with \quad \hat{P}_0 \hat{P}_N = \hat{P}_0$$

(4.2.3)

From the projection properties of the right inverse operator $\hat{N}_R^{-1}$ and $\hat{K} + \hat{G}$ it follows that the product $\hat{N}_R^{-1}(\hat{K} + \hat{G})$ is the lower triangular operator. Hence, we have a radical change in the above summation: every projection $\hat{P}_m$ of Eq.(4.2.2) contain only a finite sum of terms generated by the projection vector $\hat{P}_N \mid V >$.

## 4.3 AN EXPANSION RESULTING FROM LEFT INVERTIBILITY OF SOURCE TERM

If we use a left inverse to the lower triangular operator $\hat{G}$ related to the external forces acting on the system, we get another expansion. This left inverse satisfies equation:

$$\hat{G}_L^{-1}\hat{G} = \hat{I}$$

(4.3.1)

and it is upper triangular operator

$$\hat{G}_L^{-1} = \int \frac{\chi(\tilde{y})}{G(\tilde{y})} \hat{\eta}(\tilde{y}) d\tilde{y}$$

(4.3.2)

with an arbitrary function $\chi$ satisfying

$$\int \chi(\tilde{y}) d\tilde{y} = 1$$

(4.3.3)

Multiplying Eq.(3.2.1) by the left inverse $\hat{G}_L^{-1}$ (this is not an invertible operation), we get

$$\left(\hat{I} + \hat{G}_L^{-1}(\hat{K} + \hat{N})\right) \mid V >= 0$$

(4.3.4)

One can see that the operator $\hat{G}_L^{-1}\hat{K}$ is the right invertible operator with a right inverse $\hat{K}_R^{-1}\hat{G}$ :

$$\hat{G}_L^{-1}\hat{K} \cdot \hat{K}_R^{-1}\hat{G} = \hat{I} - \hat{P}_0$$

(4.3.5)

Multiplying (4.3.4) by a right inverse $\hat{K}_R^{-1}\hat{G}$, we get

$$\left(\hat{I} + \hat{K}_R^{-1}(\hat{G} + \hat{Q}_G\hat{N})\right)|V> = \hat{P}_{G_L^{-1}K}|V>$$

(4.3.6)

where projectors

$$\hat{Q}_G = \hat{G}\hat{G}_L^{-1}, \quad \hat{P}_{G_L^{-1}K} = \hat{I} - \hat{K}_R^{-1}\hat{G} \cdot \hat{G}_L^{-1}\hat{K}$$

(4.3.7)

The Eq. (4.3.6) can be used in the case in which the external and nonlinear forces can be treated as small perturbations to the liner part of the theory.

Substituting (4.2.2) into the l.h.s. of Eq.(4.3.6), we get a new equation

$$\left(\hat{I} + \hat{K}_R^{-1}(\hat{G} + \hat{Q}_G\hat{N})\right)\left(\hat{I} + \hat{N}_R^{-1}[\hat{K} + \hat{G}]\right)^{-1}\hat{P}_N|V> = \hat{P}_{G_L^{-1}K}|V>$$

(4.3.8)

upon the projection $\hat{P}_N|V>$. This is a **closed equation** because term causing branching

$$\hat{K}_R^{-1}\hat{Q}_G\hat{N}\hat{P}_N|V> = 0$$

(4.3.9)

The projector $\hat{P}_N$ projects on the null space of the operator $\hat{N}$. It is important that the kernels of these closed equations are explicitly given. Because operators $\hat{G}, N_R^{-1}$ are lower triangular and operators $\hat{K}_R^{-1}, \hat{K}$ are diagonals with respect to projectors (3.3.2), Eq. (4.3.8) allows us to express the projection $(\hat{I} - \hat{K}_R^{-1}\hat{Q}_G\hat{K})\hat{P}_N|V>$ of the vector $\hat{P}_N|V>$ by the lower order n-pfs generated by this vector. The crucial here are the properties of the operator

$$\hat{P}_N\left(\hat{I} - \hat{K}_R^{-1}\hat{Q}_G\hat{K}\right)\hat{P}_N$$

(4.3.10)

which, in general, is not a projector and which in the above formulation is responsible for unique solutions to projected Eq. (4.3.8):

$$\hat{P}_N\left(\hat{I} + \hat{K}_R^{-1}(\hat{G} + \hat{Q}_G\hat{N})\right)\left(\hat{I} + \hat{N}_R^{-1}[\hat{K} + \hat{G}]\right)^{-1}\hat{P}_N|V> = \hat{P}_N\hat{P}_{G_L^{-1}K}|V>$$

(4.3.11)

Now, we assume

$$\hat{P}_N\hat{P}_{G_L^{-1}K}|V> = \hat{P}_N\hat{P}_{G_L^{-1}K}|V>^{(0)}$$

(4.3.12)

where the vector $|V>^{(0)}$ corresponds to $\hat{N} = 0$. This is a weaker assumption then

$$\hat{P}_{G_L^{-1}K}|V> = \hat{P}_{G_L^{-1}K}|V>^{(0)}$$

(4.3.13)

If we take into account that the generating vector |V> is permutation symmetric, (7.16), then (4.3.6) can be substituted by

$$\left(\hat{I} + \hat{S}\hat{K}_R^{-1}(\hat{G} + \hat{Q}_G\hat{N})\right)|V> = \hat{S}\hat{P}_{G_L^{-1}K}|V>$$

(4.3.14)

and in result, the assumption (4.3.12) can be substituted by even weaker assumption

$$\hat{P}_N\hat{S}\hat{P}_{G_L^{-1}K}|V> = \hat{P}_N\hat{S}\hat{P}_{G_L^{-1}K}|V>^{(0)}$$

(4.3.15)

## 5. A RIGHT INVERSE TO $\hat{K} + \hat{G}$

Assuming that a right inverse (Green's function) to the kernel function $K$ entering the diagonal operator $\hat{K}$ (3.2.2) exists, one can check, using Cuntz co-relations (3.1.2) that the operator

$$\hat{K}_R^{-1} \equiv \int \hat{\eta}*(\tilde{z}) K_R^{-1}(\tilde{z}, \tilde{w}) \hat{\eta}(\tilde{w}) \, d\tilde{z} \, d\tilde{w}$$

(5.1.1)

satisfies equation

$$\hat{K}\hat{K}_R^{-1} = \hat{I} - \hat{P}_0$$

(5.1.2)

Now we consider equation

$$\left(\hat{K} + \hat{G}\right) \cdot \left(\hat{K} + \hat{G}\right)_R^{-1} = \hat{I} - \hat{P}_0$$

(5.1.3)

for a right inverse $(\hat{K} + \hat{G})_R^{-1}$. Multiplying (5.1.3) by a right inverse $\hat{K}_R^{-1}$ and introducing a projector

$$\hat{P}_K = \hat{I} - \hat{K}_R^{-1}\hat{K}$$

(5.1.4)

on the null space of the operator $\hat{K}$, we get

$$\left(\hat{I} + \hat{K}_R^{-1}\hat{G}\right)\left(\hat{K} + \hat{G}\right)_R^{-1} = \hat{K}_R^{-1} + \hat{P}_K\left(\hat{K} + \hat{G}\right)_R^{-1}$$

(5.1.5)

and

$$\left(\hat{K} + \hat{G}\right)_R^{-1} = \left(\hat{I} + \hat{K}_R^{-1}\hat{G}\right)^{-1}\left[\hat{K}_R^{-1} + \hat{P}_K\left(\hat{K} + \hat{G}\right)_R^{-1}\right]$$

(5.1.6)

In the above steps all operations are invertible, moreover

$$\hat{P}_K\hat{K}_R^{-1} = \hat{0},$$

(5.1.7)

The inverse in the r.h.s. of (5.1.6) exists (inverse of the unit plus lower triangular operator). Hence, it results that (5.1.6) with an **arbitrary** projection $\hat{P}_K(\hat{K} + \hat{G})_R^{-1}$ is a general solution to Eq.(5.1.3)∎

From definition, the projector

$$\hat{P}_{K+G} = \hat{I} - \left(\hat{K} + \hat{G}\right)_R^{-1}\left(\hat{K} + \hat{G}\right)$$

(5.1.8)

is a projector on the null space of the operator $\hat{K} + \hat{G}$ in space $F$, Przeworska-Rolewicz (1988). With the help of (5.1.6), one can transform (5.1.8) as follows

$$\hat{P}_{K+G} = \hat{I} - \left(\hat{I} + \hat{K}_R^{-1}\hat{G}\right)^{-1}\left[\hat{K}_R^{-1} + \hat{P}_K\left(\hat{K} + \hat{G}\right)_R^{-1}\right]\left(\hat{K} + \hat{G}\right) =$$

$$\hat{I} - \left(\hat{I} + \hat{K}_R^{-1}\hat{G}\right)^{-1}\left[\hat{Q}_K + \hat{K}_R^{-1}\hat{G} + \hat{P}_K\hat{Q}_{K+G}\right] = \hat{I} - \left(\hat{I} + \hat{K}_R^{-1}\hat{G}\right)^{-1}\left[\hat{Q}_K + \hat{K}_R^{-1}\hat{G} + \hat{P}_K\left(\hat{I} - \hat{P}_{K+G}\right)\right] =$$

$$\left(\hat{I} + \hat{K}_R^{-1}\hat{G}\right)^{-1}\hat{P}_K\hat{P}_{K+G}$$

(5.1.9)

where projectors

$$\hat{Q}_K = \hat{K}_R^{-1}\hat{K}, \quad \hat{Q}_{K+G} = (\hat{K} + \hat{G})_R^{-1}(\hat{K} + \hat{G})$$

(5.1.10)

The product $\hat{K}_R^{-1}\hat{G}$ is lower triangular operator. The last equality in (5.1.9) means that the null space of the operator $\hat{K} + \hat{G}$ is invariant with respect to the operator $\left(\hat{I} + \hat{K}_R^{-1}\hat{G}\right)^{-1}\hat{P}_K$.

# 6. A CONSTRUCTION OF A RIGHT INVERSE OPERATOR TO THE $\hat{N}(q)$

We believe that in the case of many-particle (constituent) systems, a first step to accustom non-linearity and facilitate the multi-scale description is to construct a right inverse to the operator $\hat{N}$ given by (3.2.3) or its deformed version denoted by $\hat{N}(q)$ and defined below. We do this in the case of nonlocal polynomial and local rational interactions.

## 6.1 POLYNOMIAL APPROXIMATION

Because, in general, relations between $\mu$-world and $M$-world of many particle systems, which are not reduced to special choice of initial and boundary conditions, are coded in the nonlinear terms of considered dynamic equations, we choose for the functionals $M$ and $M_q$ the most simple, namely - polynomial functionals. In the case of $q$-deformed functional $M$, (2.2.2), we have

$$M_q[\Phi; x] = \sum_n^f M_{2n}[\Phi; x, q]$$

(6.1.1)

where the $2n$-th order homogenous terms

$$M_{2n}[\lambda\Phi; x, q] = \lambda^{2n}M_{2n}[\Phi; x, q]$$

(6.1.2)

with increasing $n$ describe higher and higher nonlinear terms of interaction. Every n-order functional $M_{2n}$ can be subsequently expanded with respect to the powers of the deformation parameter $q$

$$M_{2n}[\Phi; x, q] = \sum_{k=0}^{2n} q^k M_{2n}^{(k)}[\Phi; x]$$

(6.1.3)

in which the monomial $M_{2n}^{(0)}[\Phi; x]$ describes a self-interaction of particles and the monomials $M_{2n}^{(k)}$, with k>0, are related to an interaction with other particles. The highest order monomials $M_{2f}^{(k)}$ with $k = 0, 1, ..., f$ are most responsible for the closure problem.

## 6.2 A RIGHT INVERSE TO THE OPERATOR $\hat{N}(0)$ IN CASE OF NONLOCAL $\Phi^3$-MODEL

In this case, from (3.2.3) and definitions (6.1.1-3) we have

$$\hat{N}(q) \equiv \int dz (\hat{\eta}*(z) : M_q[\hat{\eta}; z] :) = \int dz dy M(z; y)(\hat{\eta}*(\tilde{z}) : \hat{\eta}(z)(\hat{\eta}(z) - q\hat{\eta}(y))^2 :)$$

(6.2.1)

Hence

$$\hat{N}(q) = \hat{N}(0) + \hat{W}(q); \quad where \ \hat{W}(0) = \hat{0}$$

(6.2.2)

First, we construct a right inverse to the operator $\hat{N}(0)$ which describe self interaction of particles and which, for small $q$, is most important in the above expansion. This operator, in the case of polynomial version of the formula (2.2.2) is given by

$$\hat{N}(0) = \int dz dy M(z; y)(\hat{\eta}*(z)\hat{\eta}(z))(\hat{\eta}(z))^2 \equiv \int dz M(z)(\hat{\eta}*(z)\hat{\eta}(z))(\hat{\eta}(z))^2$$

(6.2.3)

with

$$M(z) = \int dy M(z; y)$$

We used the same notation as in (2.1.4) where

$$\hat{\eta}(\tilde{x}) = \hat{\eta}(\alpha, \vec{x}, t)$$

(6.2.4)

are generators of the C*-algebra (one type of particles). Lack of tilde over the variable $z$ means that quantities appearing in the parenthesis are summed up with respect to discrete variable $\alpha$. $\hat{N}(0)$ is a local operator. We construct a right inverse $\hat{R}(0)$ to the operator (6.2.3), which from definition satisfies equation

$$\hat{N}(0)\hat{R}(0) = \hat{I} - \hat{P}_0$$

(6.2.5)

where $\hat{R}(0) \equiv (\hat{N}(0))_R^{-1}$ in the notation used in Sec.4. A possible solution is

$$\hat{R}(0) = 3^{-1} \int dy (\hat{\eta}*(y))^2 1/M(y)$$

(6.2.6)

what can be seen by substitution (we do not need to afraid zeros of function $M(y)$ which is a constant. Usually, $M(y, z) = M(y - z)$ ). For a proof of (6.2.5), we assume that all variables are discrete and the integration is substituted by the summation. We have used the following property resulting from the Cuntz relations (3.1.2) and definition (2.1.4):

$$(\hat{\eta}(z))^2 \cdot (\hat{\eta}*(y))^2 = \sum_{\alpha, \beta} \hat{\eta}^2(\alpha, z) \cdot \hat{\eta}^{*2}(\beta, y) = \sum_{\alpha, \beta} \delta_{\alpha\beta}\delta_{\alpha\beta}\delta_{zy}\delta_{zy}\hat{I} = 3\delta_{zy}\hat{I}$$

(6.2.7)

where $\delta$ designates many dimensional Kronecker deltas.

Another solution to Eq. (6.2.5) is given by

$$\hat{R}(0) = 3^{-1} \int dy (\hat{\eta} * (y))^2 (1/M(y))(\hat{\eta} * (y)\hat{\eta}(y))$$

(6.2.8)

For a general solution, see Przeworska-Rolewicz (1988) or I.

A projector on the null space of the operator $\hat{N}(0)$ is then

$$\hat{P}_{N(0)} = \hat{I} - \hat{R}(0)\hat{N}(0) \equiv \hat{I} - \hat{Q}_{N(0)}$$

(6.2.9)

where in case (6.2.6), the projector

$$\hat{Q}_{N(0)} = \hat{R}(0)\hat{N}(0) = 3^{-1} \int d\tilde{y} d\tilde{z} \frac{M(\tilde{z})}{M(\tilde{y})} \hat{\eta} *^2 (\tilde{y}) \hat{\eta} * (\tilde{z}) \hat{\eta}(\tilde{z}) \hat{\eta}^2(z)$$

(6.2.10)

where integration or summation is understood over all variables under integration sign and the convention (2.1.4) is used. See also (6.2.4).

## 6.3 A RIGHT INVERSE TO $\hat{N}(q)$ IN $\Phi^3$-CASE

We consider equation

$$\hat{N}(q)\hat{R}(q) = \hat{I} - \hat{P}_0$$

(6.3.1)

and we look for a right inverse to the operator $\hat{N}(q)$ in the form

$$\hat{R}(q) = \hat{R}(0)\hat{Y}(q)$$

(6.3.2)

First, we calculate

$$\hat{N}(q)\hat{R}(0) = \left(\hat{N}(0) + q\hat{N}_1 + q^2\hat{N}_2\right)\hat{R}(0)$$

where the operator $\hat{R}(0)$ is given e.g. by (6.2.6). We get

$$\hat{N}(q)\hat{R}(0) = \left(\hat{N}(0) + q\hat{N}_1 + q^2\hat{N}_2\right)\hat{R}(0) =$$
$$\int dz dy dx M(z;y)\{(\hat{\eta} * (z)\hat{\eta}(z))[(\hat{\eta}(z))^2 - 2q(\hat{\eta}(z)\hat{\eta}(y)) + q^2(\hat{\eta}(y))^2]\}3^{-1}(\hat{\eta} * (x))^2 1/M(x) =$$
$$\hat{I} - \hat{P}_0 + \int dz \left(-2q\frac{M(z;z)}{M(z)} + q^2 \int dy \frac{M(z;y)}{M(y)}\right)(\hat{\eta} * (z)\hat{\eta}(z)) \equiv \hat{I} - \hat{P}_0 + \int dz O(z)(\hat{\eta} * (z)\hat{\eta}(z))$$

(6.3.3)

where the function

$$O(z) = -2q\frac{M(z;z)}{M(z)} + q^2 \int dy \left(\frac{M(z;y)}{M(y)}\right)$$

(6.3.4)

see (6.2.3). Hence, in the case of (6.2.6), the operator $\hat{Y}$ in (6.3.2) is

$$\hat{Y} = \int dz \left(\frac{1}{1 + O(z)}(\hat{\eta} * (z)\hat{\eta}(z))\right)$$

and

$$\hat{R}(q) = 3^{-1} \int dy\, dz \left( \frac{1}{M(y)(1 + O(z))} (\hat{\eta}*(y))^2 (\hat{\eta}*(z)\hat{\eta}(z)) \right)$$

(6.3.5)

## 7. NON-POLYNOMIAL INTERACTION.

In the Newton theory, the "field" $\Phi(\tilde{x})$ at the "point" $\tilde{x}$, and at the moment $t$, in Eq.(2.1.1), depends on values of the field at other points at the same time $t$ (action at a distance). In fact, at derivation of continuum mechanic equations - the non-locality - plays less role due to coarse graining (Navier-Stokes equations are local again). A more basic theories (relativistic quantum field theory (RQFT)) are local again. There are arguments against non-locality, Kondepudi et al. (1999), and in RQFT an action at a distance is treated as the zeroth order approximation to interaction among charged particles, (no electromagnetic field), see *Feynman Lectures on Gravitation*.

In this section we assume a local nonlinear theory (2.1.1). Non-locality is admitted only in the linear term of Eq. (2.1.1). In particular, in (2.1.3) and (3.2.3) there are only variables $\Phi(\tilde{x})$ and $\hat{\eta}(\tilde{z})$. In fact sometimes in literature, dynamics with infinitely many derivatives are considered as nonlocal, Barnaby et al. (2008). From that point of view, considered here rational interactions can be named as nonlocal (in the symmetrical Fock space the operators $\hat{\eta}(\tilde{x})$ are represented by the functional derivatives, Rzewuski (1969)).

Second departure from assumptions used up to now concerns a polynomial restriction. This assumption is justified by a mathematical simplicity of such theories like renormalization of perturbation expansions in quantum field theory or by small values of considered variables like in the case of displacements of solid state particles. Now we consider a richer class of interactions - some times named as the *essential nonlinear interactions* described, for example, by rational functions. In spite of the fact that it is easy to give an example of non-polynomial theory for which Eq.(3.2.1) can be transformed in equation preserving all properties of the polynomial theories, they posses a new interesting properties which can be useful in practical calculations. From that point of view would be interesting to know all *equivalent systems of equations,* which from definition have the same set of solutions, Lopuszanski (1998).

First, we consider the following equation

$$\left( \hat{K} + \hat{G} + \lambda \left( \hat{I} - \hat{N}_{loc} \right)^{-1} \right) | V >= 0$$

(7.1)

where the operator $\hat{N}_{loc}$ represents a local interaction considered before and in I, and is a right invertible operator. If in this equation the inverse admits an expansion

$$\left( \hat{I} - \hat{N}_{loc} \right)^{-1} = \sum (\hat{N}_{loc})^n$$

(7.2)

then it is easy to see that every power in the r.h.s. of (7.2) is again a local expression. Multiplying (7.1) by $(\hat{I} - \hat{N}_{loc})$ we get equation

$$\left( (\hat{I} - \hat{N}_{loc})(\hat{K} + \hat{G}) + \lambda \hat{I} \,|\, V > \right) =$$
$$\left( \lambda \hat{I} + \hat{K} + \hat{G} - \hat{N}_{loc}(\hat{K} + \hat{G}) \right) |V >= 0$$

(7.3)

in which $(\lambda \hat{I} + \hat{K})$ is diagonal, $\hat{G}$ is lower triangular, $\hat{N}_{loc}\hat{K}$ is a right invertible if $\hat{N}_{loc}$ and $\hat{K}$ are such and $\hat{N}_{loc}\hat{G}$ is a diagonal or diagonal + upper triangular operators.

**It is worth to notice** that an expansion of the vector |V> in the positive powers of the operator $\hat{N}_{loc}$, see (4.1.4), is possible only in the case of singular operator $(\lambda \hat{I} + \hat{K})$ (in the case of Eq.(3.2.1) we need a singularity of the operator $\hat{K}$ ). This means that a possibility of expansion of the vector |V> in positive powers of a more elementary local interaction $\hat{N}_{loc}$ needs special tuning of the values of the coupling constant $\lambda$, namely, they have to be eigenvalues of the operator $\hat{K}$.

If the above assumption is not fulfilled, we can use expansions based on the right invertibility of the product $\hat{N}_{loc}\hat{K}$. From (7.3), we can get

$$\left( \hat{I} - (\hat{N}_{loc}(\hat{K} + \hat{G}))_R^{-1}(\lambda\hat{I} + \hat{K} + \hat{G}) \right) | V > = \hat{P} | V >$$

(7.4)

with projectors

$$\hat{P} = \hat{I} - (\hat{N}_{loc}(\hat{K} + \hat{G}))_R^{-1} \hat{N}_{loc}(\hat{K} + \hat{G}) = \hat{I} - (\hat{K} + \hat{G})_R^{-1}\hat{Q}_{loc}(\hat{K} + \hat{G}) \quad and \quad \hat{Q}_{loc} \equiv (\hat{N}_{loc})_R^{-1}\hat{N}_{loc}$$

(7.5)

where we took into account

$$(\hat{N}_{loc}(\hat{K} + \hat{G}))_R^{-1} = (\hat{K} + \hat{G})_R^{-1}(\hat{N}_{loc})_R^{-1}.$$

(7.6)

(7.4) is an interesting equation since the coupling constant $\lambda$ stands in the front of the lower triangular operator $(\hat{N}_{loc}(\hat{K} + \hat{G}))_R^{-1}$. This means that if the **arbitrary** projection $\hat{P} | V >$ **does not depend on the** $\lambda$, we get, for the generating vector $| V >$, finite, exact series in the positive powers of $\lambda$. This means that, for the essentially nonlinear theory (7.1) of the rational type, finite series in the positive powers of the coupling constant $\lambda$ satisfy exact equations for correlation functions. Moreover, choosing

$$\hat{P} | V > = \hat{P} | V >^{(0)} = | V >^{(0)}$$

(7.7)

we find that

$$\lim_{\lambda \to 0} | V > \to | V >^{(0)}$$

(7.8)

However, the choice (7.7) can be in a contradiction with permutation symmetry of correlation functions. So, at the end of paper we give a weaker restriction of the arbitrary projection $\hat{P} | V >$ then (7.7), namely (7.18), leading automatically to symmetrical solutions. For comparison, see (4.3.15).

A generalization of Eq. (7.1) can be done as follows

$$\left( \hat{K} + \hat{G} + \lambda(\hat{I} - \hat{N}_{loc})^{-1}\hat{M}_{loc} \right) | V > = 0$$

(7.9)

Hence,

$$\left( (\hat{I} - \hat{N}_{loc})(\hat{K} + \hat{G}) + \lambda\hat{M}_{loc} \right) | V > = 0$$

(7.10)

After similar transformations as in Eq. (7.1), these equations lead again to exact solutions in the finite powers of the coupling constant $\lambda$, if the right invertible operator $\hat{N}_{loc}$ has a bigger upper triangular order then $\hat{M}_{loc}$. In the case of the same order, we get from (7.10) closed equations. (show this). It looks as if interactions (self or not, or local or non-local) of the type $\lambda(\hat{I} - \hat{N}_{loc})^{-1}$ or $\lambda(\hat{I} - \hat{N}_{loc})^{-1}\hat{M}_{loc}$ have great power of formal stability; only a few terms in positive powers of the coupling constant $\lambda$ are needed to describe exact solutions. But a few words of justification is required. Can we expect that inverses appearing in Eqs (7.1) and (7.9) exists? In the Fock space $F$ considered in I and here one can construct right inverses to the operator $(\hat{I} - \hat{N}_{loc})$ and then Eqs (7.3) and (7.10) can be also obtained. We have assumed non-singular character of the operator $(\hat{I} - \hat{N}_{loc})$ to "justify", via (7.2), local character of considered non-polynomial interaction. In fact, for a small $\hat{N}_{loc}$ the above assumption is plausible. If, however, the operator $(\hat{I} - \hat{N}_{loc})^{-1}$ is substituted by a right inverse $(\hat{I} - \hat{N}_{loc})^{-1}_R$, then corresponding passages from (7.1) to (7.3) and from (7.9) to (7.10) are not equivalent. The obtained equations are more general then primary ones. In other words, like in the renormalization theory we execute some operations upon formal quantities (divergent integrals) to give them a sense in the frame of renormalization theory. In some way we can also look on the equations with rational interactions as on equations with infinitely many derivatives with corresponding *initial value problem*, see Barnaby et al. (2008).

At the end we transform Eq.(7.9) to a form useful for further considerations:

$$\left(\hat{I} + \lambda\hat{R}\hat{M}_{loc}\right)| V \rangle = \hat{P} | V \rangle$$

(7.11)

where a right inverse operator $\hat{R}$ satisfies equation

$$\left(\hat{I} - \hat{N}_{loc}\right)\left(\hat{K} + \hat{G}\right) \cdot \hat{R} = \hat{I}$$

(7.12)

This operator we can look for in a form

$$\hat{R} = \left(\hat{K} + \hat{G}\right)^{-1}_R \left(\hat{N}_{loc}\right)^{-1}_R \hat{Y}$$

(7.13)

what leads to a simple equation

$$\left((\hat{N}_{loc})^{-1}_R - \hat{I}\right)\hat{Y} = \hat{I}$$

(7.14)

with a lower triangular operator $(\hat{N}_{loc})^{-1}_R$. The projector

$$\hat{P} \equiv \hat{I} - \hat{R} \cdot \left(\hat{I} - \hat{N}_{loc}\right)\left(\hat{K} + \hat{G}\right)$$

(7.15)

and the projection $\hat{P} | V \rangle$ appearing in (7.13) is an arbitrary vector from the null space of the operator $\left(\hat{I} - \hat{N}_{loc}\right)\left(\hat{K} + \hat{G}\right)$. Because we are interested in symmetric solutions

$$| V \rangle = \hat{S} | V \rangle$$

(7.16)

see I, we consider projected (symmetrized) Eq.(7.11)

$$\left( \hat{I} + \lambda \hat{S} \hat{R} \hat{M}_{loc} \right) | V >= \hat{S} \hat{P} | V >$$

(7.17)

with

$$\hat{S} \hat{P} | V >= | V >^{(0)}$$

(7.18)

where the vector $| V >^{(0)}$ is a symmetrical vector satisfying (7.18) with the coupling constant $\lambda = 0$. (7.18) can be interpreted as an restriction of a freedom of the perturbation theory, see Bogolubov et al. (1976) and I. In (7.18) we also assume that an arbitrary element of the solution to Eq.(7.9), the projection $\hat{P} | V >$, can be chosen in such a way that its symmetrical projection, the vector $\hat{S} \hat{P} | V >$ does not depend on the coupling constant $\lambda$.

It turns out that equations satisfied by the Green's functions of quantum field theory (QFT), for non-polynomial (rational) interactions, have similar structure as Eq. (7.10). In that case the role of operator $\hat{G}$ is played by the source operator which is also a lower triangular with respect to projectors (3.3.1-2) and left invertible operator, Hańćkowiak (1979). Hence, it results that in QFT similar equations to Eqs (7.10) can be used for a construction of solutions tending in the limes $\lambda \to 0$ to the free solutions described by an equation with similar structure as Eq.(7.10). It is interesting that due to presence of a right inverse operator $\hat{R}$, Eqs (7.17) can be divided in three classes according to projection properties of the product $\hat{R} \hat{M}_{loc}$, namely this product can be a diagonal, lower triangular or upper triangular operator with respect to projectors $\hat{P}_n$ upon n-point functions, see I. In first two cases, we can say that for definite class of rational interactions, it is possible to get closed equations for correlation functions.

# 8. FINAL REMARKS

## 8.1 COARSE AND FINE GRAIN DESCRIPTION

Different levels of descriptions of n-particle systems are realized by different averages denoted usually by $< \dots >$. Using apparatus with appropriate resolutions (averages) and appropriate forces, patterns can emerge. We believe that the free Fock space, with plenty easily invertible operators, will help to see this phenomenon and simultaneously allows us to develop a multiscale description of different parts of the system, Solomon (1996), El-Azab (2004), Liu et al. (2004), Wijessinghe et al. (2004).

## 8.2 INTERPRETATION OF AVERAGES

Now let us say a few words about an interpretation of the above quantities like $< \Phi [\tilde{x}, t; \bullet] >$. There are at least three reasons to use averaged (smeared) quantities: first we do not control initial conditions with micro-precision, second, - precise, even macro-solutions, are often hard to obtained, third, - solutions are very sensitive with respect to small perturbations and changes of initial conditions - and only averaged solutions can be easier approximated and can be confronted with measurements.

Very often, after identification of areas (particles) requiring a more precise description (cracks, dislocations, interfaces, routers) we relax, from economical reasons, a precision of description for the rest of the system.

In the case in which $< \Phi [\tilde{x}; \bullet] >$ suggests the crack in some area, the following alternative arises: all solutions of the ensemble correspond to the crack or solutions of the ensemble cancel each other in the crack area. In fact, we can resolve the above uncertainty by analysing the 2-point correlation function.

By separating of a number of variables from a number of degrees of freedom, the reversible micro-models with many degrees of freedom and related number of variables are gradually substituted by "hydrodynamic" models with a smaller number of freedoms but with the same number of variables. This preserve essential feature of formalism – an reversibility of used operators - in spite of averaging operations. A reduction of number of freedoms in that case is reflected by similarity of all these variables. In the most extreme case, the macro system can be described by six different sets of identical variables (rigid body). In fact, the above averaging operations are similar to an identification of points with the help of some equivalent relation: equivalent classes are described by the same number of points, but many of them are equal to each others.

## 8.3 AVERAGES OVER WHOLE SPACE-TIME

In the case of smearing over whole space-time in formulas (1.2.7), what corresponds to a weight function

$$W = const$$

(8.3.1)

(no macroscopic constraints but only Laplace's principle of equal ignorance), all coarse grained 1-point variables are identical. In this case the two point correlation functions can be identified with autocorrelation functions. It is astonishing that in this case, by means of the convolution theorem, an exact solution can be retrieved from them. From the convolution theory we have

$$\mathscr{F}[f * g] = \mathscr{F}[f]\mathscr{F}[g],$$

(8.3.2)

where the convolution

$$(f * g)(t) = \int_{-\infty}^{\infty} g(t')f(t-t')dt'$$

(8.3.3)

contains integration over the whole $R$. For $f \equiv g$, the convolution (8.3.3) can be identified with the 2-point autocorrelation function considered in (1.2.7) with weight function (8.3.1). Can we uniquely retrieve a function from its square? Sometimes yes sometimes no. For example, knowing $y = \sin^2 x$ and taking square root from $y$ we can calculate $\sin x$ and in places where $y$ is not differentiable, we can reconstruct its further behaviour changing its sign.

## 8.4 HYDRODYNAMIC DESCRIPTIONS

To describe Hydrodynamic using statistical mechanics, we need local quantities, Resibois et al. (1977), like the *density field*

$$n(\vec{r},t) = \int f_1(\vec{r},\vec{v},t)d\vec{v}$$

(8.4.1)

the *velocity field*

$$\vec{u}(\vec{r},t) = \frac{1}{n(\vec{r},t)}\int \vec{v}f_1(\vec{r},\vec{v},t)d\vec{v}$$

(8.4.2)

and the *temperature field*

$$T(\vec{r},t) = \frac{1}{d\,k_B n(\vec{r},t)} \int \frac{mv^2}{2} f_1(\vec{r},\vec{v},t)d\vec{v}$$

(8.4.3)

where $f_1$ is the Boltzmann function defined by (1.1.3-4). We would like to see how MTCF (1.2.1) and (1.2.7) or (2.4.1) are related to the above quantities. Let us concentrate upon the case of incompressible fluid with constant temperature. So we assume that $n$ and $T$ are constants and only the velocity field $u$ given by (8.4.2) and satisfying the Navier-Stokes equation is considered. Let us introduce the moments

$$\vec{u}(t)^{klm} \equiv \int x_1^k\, y_1^l\, z_1^m\, \vec{u}(\vec{r}_1,t) n(\vec{r}_1,t) d\vec{r}_1 = \int x_1^k\, y_1^l\, z_1^m\, \vec{u}_1 f_1(\vec{r}_1,\vec{u}_1,t) d\vec{r}_1 d\vec{u}_1$$

(8.4.4)

where $\vec{r}_1 = (x_1, y_1, z_1)$. We would like to see how these moments are related to MTCF (1.2.1) or (1.2.7) ((2.4.1)). We have

$$\vec{u}(t)^{klm} \equiv \int x_1^k\, y_1^l\, z_1^m\, \vec{u}(\vec{r}_1,t) f(\vec{r}_1,\vec{u}_1,...,\vec{r}_N,\vec{u}_N,t) d\vec{r}_{(N)} d\vec{u}_{(N)}$$

(8.4.5)

where 6N point function $f$ is the positive integral of motion of Eq.(1.1), which satisfies the Liouville's equation (1.1.1). Introducing new variables $Y$ in the integral (8.4.5)

$$X \equiv (\vec{r}_1,\vec{u}_1,...,\vec{r}_N,\vec{u}_N) = X(t;Y)$$

(8.4.6)

such that $X$ satisfies Eq.(1.1), we can eliminate the $t$-dependence from the function $f$. If additionally, the Jacobian of these one parameter family of transformations is equal to one (Liouville's theorem), we get exactly that

$$\vec{u}(t)^{klm} = \int \Phi^k[1,\vec{r}_1,t;Y]\Phi^l[2,\vec{r}_1,t;Y]\Phi^m[3,\vec{r}_1,t;Y]\dot{\vec{\Phi}}[\vec{r}_1,t;Y]f[Y]d^{6N}Y$$

(8.4.7)

where notation (1.2.5) was used and $k,l,m = 0,1,2....$ Hence, for incompressible fluid, for which the number of particles in unit volume, $n$, is constant, moments (8.4.4) of hydrodynamic velocity (8.4.2) are proportional to the MTCF (1.2.1) with equal times and equal radius vectors $\vec{r}_1$. We remind you that in notation (1.2.5), the radius vector in the field $\Phi$ identifies a particle. So, with the help of MTCF (1.2.1) one can retrieve the basic hydrodynamic quantity of incompressible fluid, see also Huang (1963).

## 8.5 A ROLE OF AVERAGES

We propose the following description of the classical world or, at least, its part: we start from an equation like (2.1.1) describing the whole world with all particularities admitted by physical laws. A particular system, e.g., the solar sytem, can be considered as an appropriately averaged original world in a scale in which planets are described by points. In fact, in this way Mach was looking on the Galilelian inertial principle.

By averaging procedures, we may reduce number of degrees of freedom (not variables) up to our will, for example, to one degree of freedom (one super particle (world)). A choice of the function $W$ in (1.2.7) depends on chosen scale(s) related to used instruments at different times. Choice of the function $f$ in (2.4.1) is related to accepted initial macro conditions which can be described by the Boltzmann-Gibbs approach. If we are able to control more precisely some constituents of the system, the form (2.4.2) for $f$, can be used. In this case, Eq.(2.1.1) corresponds rather but not necessary (progress in atom manipulations) to some averaged original particles (subsystem). In presented description of the classical world, the resulting force acting on a "point" is given by

$$\mu < \ddot{\Phi}(\tilde{x}) >$$

(8.5.1)

where $\mu$ is the mass of an extracted subsystem and $<...>$ describes averaging related to a chosen scale.

In principle, systems driven away from an initial state of thermal equilibrium, what is accomplished by external forces representing moving walls, pistons and so on, can be described by the developed formalism.

## 8.6 A GENERALIZED INVERSE $G$

to a given operator $A$ is defined by *general condition*

$$\hat{A}\hat{G}\hat{A} = \hat{A}$$

(8.6.1)

It is interesting to notice that a right inverse to the operator $\hat{A}$, see, e.g., (4.1.1), and denoted in paper by $\hat{A}_R^{-1} \equiv \hat{G}$, satisfies this condition. The same is true to a left inverse operator. Moreover, the operators

$$\hat{Q} = \hat{G}\hat{A} = \hat{Q}^2, \quad \hat{Q}' = \hat{A}\hat{G} = \hat{Q}'^2$$

(8.6.2)

are projectors (idempotent) . This results from general condition (8.6.1) and associative property of used operators. When an operator $A$ is left or right invertible operator, then one of the above group of equations is trivial. If we know an operator $\hat{G}$ satisfying Eq.(8.6.1), then general solution to equation

$$\hat{A} | V >= 0$$

(8.6.3)

can be represented as

$$| V >= \hat{P} | V >$$

(8.6.4)

where a projector on the null space of operator $\hat{A}$

$$\hat{P} \equiv \hat{I} - \hat{Q} = \hat{I} - \hat{G}\hat{A}$$

(8.6.5)

(from (8.6.1) we have $\hat{A}\hat{P} = \hat{0}$).

The *Moore-Penrose generalized inverse* $\hat{G}$ satisfies additional three conditions:

$$\hat{G}\hat{A}\hat{G} = \hat{G} \quad \text{(reflexive condition)},$$

(8.6.6)

$$\left(\hat{A}\hat{G}\right)^* = \hat{A}\hat{G} \quad \text{(normalized condition)}$$

(8.6.7)

$$\left(\hat{G}\hat{A}\right)^* = \hat{G}\hat{A} \quad \text{(reverse normalized condition)}$$

(8.6.8)

Then, at least in finite dimensional spaces, $\hat{G}$ matrices are unique and corresponding solutions to finite dimensional equations (8.6.3) have the **smallest Euclidean norms**. For right inverse operators, only the last condition is not satisfied. So, we cannot expect minimal norm solutions, which, in our case, have no physical meaning. Nevertheless, the Moore-Penrose inverses have interesting applications in so called frame expansions, see Christensen (1994, 1999).

A *frame* is a family $\{f_i\}_{i \in I}$ of elements of a given linear space $F$ with the property that every element in the space can be written as a (infinite) linear combination of the frame elements. The motivation behind frames is that a frame $\{f_i\}_{i \in I}$ can be over completed, so there is freedom in the choice of the expansion coefficients. In paper the above property was used for construction different right and left inverses operators.

If we postulate conditions (8.6.1), (8.6.6) and

$$\hat{A}\hat{G} = \hat{G}\hat{A}$$

(8.6.7)

then the unique operator $\hat{G}$ satisfying these three conditions is called the *group inverse* and has application in the problems on optimisation solution of differential equations. For a realization of such inverses see Cao (2006).